\documentstyle[aps,prd,psfig]{revtex}
\begin{document}
\renewcommand{\thesection}{\arabic{section}}
\renewcommand{\thesubsection}{\arabic{subsection}}

\title{Relativistic Version of Landau Theory of Fermi Liquid in Presence
of Strong Quantizing magnetic Field- An Exact Formalism}

\author{
Sutapa Ghosh$^{a)}$, Soma Mandal$^{a)}$ and Somenath Chakrabarty$^{a),b)}
 ${\thanks{E-Mail: somenath@klyuniv.ernet.in}}}
\address{$^{a)}$Department of Physics, University of Kalyani, Kalyani 741 235,
India and $^{b)}$Inter-University Centre for Astronomy and Astrophysics, Post 
Bag 4, Ganeshkhind, Pune 411 007, India}
\date{\today}
\maketitle
\begin{abstract}
An exact formalism for the relativistic version of Landau theory of
Fermi liquid in presence of strong quantizing magnetic field is
developed. Both direct and exchange type interactions with scalar and
vector coupling cases are considered.
\end{abstract}
\section {Introduction}
The study of dense stellar matter under ultra-strong magnetic field has
gotten a new life after the discovery of a few magnetars \cite{R1,R2,R3,R4}. 
These stellar objects are believed to be strongly magnetized young neutron 
stars. The surface magnetic fields are observed to be $\geq 10^{15}$G. Then 
it is quite possible that the field at the core region may go up to $10^{18}$G.
The exact source of this strong magnetic field is not known. These objects are
also supposed to be the possible sources of anomalous X-ray and soft gamma 
emissions (AXP and SGR). If the magnetic field is  really so strong, in 
particular at the core region, they must affect most of the important physical 
properties of such stellar objects. The elementary processes, e.g., weak and
electromagnetic processes taking place at the core region should also change
significantly.

The strong magnetic field affects the equation of state of dense neutron
star matter and as a consequence the gross-properties of neutron stars 
\cite{R5,R6,R7,R8}, e.g., mass-radius relation, moment of inertia,
rotational frequency etc. should change significantly. In the case of
compact neutron stars, the phase transition from neutron matter to quark 
matter at the core region is also affected by strong quantizing magnetic field.
It has been shown that a first order phase transition initiated by the
nucleation of quark matter droplets is absolutely forbidden if the magnetic 
field $\sim 10^{15}$G at the core region \cite{R9,R10}. However a second order 
phase transition is allowed, provided the magnetic field strength $<10^{20}$G. 
This is of course too high to achieve at the core region.

The elementary processes, in particular, the weak and the electromagnetic
processes taking place at the core region of a neutron star are strongly 
affected by such ultra-strong magnetic field \cite{R11,R12}. Since the cooling 
of neutron stars are mainly controlled by neutrino/anti-neutrino emission, the 
presence of strong quantizing magnetic field should affect the thermal history 
of strongly magnetized neutron stars.  Further, the electrical conductivity of 
neutron star matter which directly controls the evolution of neutron star 
magnetic field will also change significantly.

In another kind of work, the stability of such strongly magnetized rotating 
objects are studied. It has been observed from the detailed general 
relativistic calculation that there are possibility of some form of geometrical
deformation of these objects from their usual spherical shapes 
\cite{R13,R14,R15}. In the extreme case such objects may either become 
black strings or black disks. In the non-extreme case, however, it is
also possible to detect gravity waves from these rotating objects.

In a recent study of microscopic model of dense neutron star matter,
we have observed that if most of the electrons occupy
the zeroth Landau level, with spin anti-parallel to the direction of
magnetic field and very few of them are with spin along the direction of
magnetic field and Landau quantum number $>0$, then either such strongly
magnetized system can not exist or such a strong magnetic field can not be 
realized at the core region of a neutron star \cite{R16}. We have shown in 
that work that since the electrical conductivity of the medium becomes 
extremely low in presence of ultra-strong magnetic field, the
magnetic field at the core region must therefore decay very rapidly.
Hence we have argued that the second possibility is more feasible, i.e.,
strong magnetic field can not exist at the core of a magnetar.

So far most of the calculations on the equation of states of dense
stellar matter in presence of strong magnetic field are either based on
some kind of mean field approximation or non-relativistic potential
model \cite{R6,R17}.  In this paper we shall derive an exact formalism of 
relativistic version of Landau theory of Fermi liquid in presence of strong 
quantizing magnetic field and obtain the quasi-particle energy for both scalar 
and vector coupling cases. We shall consider a typical many body
fermionic system in presence of strong magnetic field interacting via
some kind of scalar and vector bosons exchange. Therefore from this
investigation it is very easy to obtain equation of state of a real fermion
system in presence of strong magnetic field, e.g., high density electron
gas (with photon exchange), dense neutron star matter (with
$\sigma-\omega-\rho$ meson exchange) or even dense quark matter produced
at the core region as a result of quark-hadron phase transition 
(exchange of color gluons). Not only that, one can
study the magnetic as well as the non-equilibrium  properties of such dense
fermionic system as mentioned above. The relativistic version of this
model without magnetic field was developed long ago by Baym and Chin
\cite{R18}.  So far most of the scattering
or reaction or decay rates/cross-sections are evaluated either with very low
magnetic field approximation or in presence of ultra-strong magnetic
field. To the best of our knowledge the present article is the
first attempt to develop an exact theoretical formalism to obtain the
two-body scattering matrix element (both direct and exchange) for all possible
strengths of external magnetic field. In a completely different type of
work we have used  this formalism to obtain the rates of weak and 
electro-magnetic processes. We have investigated the effect of strong magnetic 
field on the emissivities and mean free paths of neutrinos / anti-neutrinos
and electrical conductivity of dense electron gas in neutron star
matter. These quantities play significant role in thermal evolution and the 
evolution of magnetic fields of neutron stars respectively.

Now similar to the conventional non-relativistic case, the relativistic 
version also deals with normal Fermi liquid and applicable only for the 
low-lying excited states of the system, which are made of superpositions of 
quasi-particle excitations close to the Fermi surface. The relativistic version
is therefore applicable to high density electron gas, dense neutron star
matter and also to dense quark matter which may present at the core
region of compact neutron stars. For the applicability of this model the
temperature of the systems are therefore assumed to be low enough compared to 
the chemical potential of the constituents.

The paper is organized in the following manner. In the next section we shall 
develop an exact formalism for the relativistic version of Landau theory of 
Fermi liquid considering a typical many body fermionic system interacting via 
some kind of scalar and vector boson exchange field. In the last section
we have discussed the results and future prospects of the work.

\section{Formalism}
We have considered a dense relativistic fermionic system at zero
temperature ($T << \mu_f$, the chemical potential of the fermions)
and interacting via scalar and vector boson fields represented  by $\phi$
and $V^\mu$ respectively with masses $\zeta$ and $\eta$ respectively. The
corresponding coupling constants with the fermionic field are $g_s$ and
$g_v$ respectively. Then the basic Lagrangian density is given by
\begin{eqnarray}
\cal{L}&=&\frac{i}{2}\bar \psi \gamma^\mu(\vec
D_\mu-D^{\!\!\!\!\!^{^\leftarrow}}_\mu)
\psi -m\bar
\psi \psi\nonumber \\ &-&g_v\bar \psi \gamma^\mu V_\mu \psi-g_s \bar \psi
\phi \psi +{\cal{L}}_v+{\cal{L}}_s+{\cal{L}}_{{\rm{em}}}
\end{eqnarray}
Where $D^\mu=\partial^\mu+iqA^\mu$ with $q$ is the magnitude of electric
charge, $m$ is the rest mass of the fermions and $A^\mu\equiv 
(0,0,xB_m,0)$, the electro magnetic field vector. Here the gauge
choice is such that the constant external magnetic field $B_m$ is along
Z-axis, ${\cal{L}}_v$, ${\cal{L}}_s$ and ${\cal{L}}_{{\rm{em}}}$ are the free
Lagrangian densities for the vector, scalar and the electromagnetic fields
respectively. Therefore, in this particular case, ${\cal{L}}_{{\rm{em}}}$ is the
Lagrangian density corresponding to the constant magnetic field $B_m$.
We now adiabatically turn on the interaction between the fermions.
Therefore, the normal Fermi liquid system will evolve from the non-interacting
ground state into the interacting ground state and there will be a
one-to-one correspondence between the bare particle states of the
original system and the dressed or quasi-particle states of the interacting 
system.

Next to obtain $f$ the Landau Fermi liquid interaction, which is also
called the quasi-particle interaction function, we have considered both
direct and exchange type interaction. The exchanged bosons are either
scalar or vector type. The Landau Fermi liquid interaction $f$ is related
to the two-particle forward scattering amplitude. Therefore, it is
essential to compute transition matrix element for two-particle direct as
well as exchange interaction forward scattering via the exchanged
scalar and vector bosons. We further assume that the constant magnetic 
field acts as a background field and is present throughout the system. 
We further assume that the strength of magnetic field is such that in the
relativistic case the Landau levels are populated for the charged
fermions, i.e., $B_m\geq B_m^{(c)(f)}$, the quantum mechanical limit,
given by $qB_m^{(c)(f)}=m^2$.

Then  the modified form of spinor solutions of charged fermions in presence of
strong quantizing magnetic field is given by
\begin{equation}
\psi=\frac{1}{(L_yL_z)^{1/2}}\exp\{-iE_\nu t+ip_yy+ip_zz\}u^{\uparrow
\downarrow}
\end{equation}
where
\begin{equation}
u^\uparrow=\frac{1}{[2E_\nu(E_\nu+1)]^{1/2}} \left (\begin{array}{c}
(E_\nu+m)I_{\nu;p_y}(x)\\ 0\\ p_zI_{\nu;p_y}(x)\\ -i(2\nu qB_m)^{1/2}
I_{\nu-1;p_y}(x) \end{array} \right )
\end{equation}
and
\begin{equation}
u^\downarrow=\frac{1}{[2E_\nu(E_\nu+1)]^{1/2}} \left (\begin{array}{c}
0\\ (E_\nu+m)I_{\nu-1;p_y}(x)\\ i(2\nu qB_m)^{1/2} I_{\nu;p_y}(x)\\
-p_zI_{\nu-1;p_y}(x)
 \end{array} \right )
\end{equation}
where the symbols $\uparrow$ and $\downarrow$ indicates up and down spin states.
\begin{eqnarray}
I_\nu=\left (\frac{qB_m}{\pi}\right )^{1/4}\frac{1}{(\nu !)^{1/2}}2^{-\nu/2}
\exp \left [{-\frac{1}{2}qB_m\left (x-\frac{p_y}{qB_m} \right )^2}\right 
]\nonumber \\
H_\nu \left [(qB_m)^{1/2}\left (x-\frac{p_y}{qB_m} \right) \right ]
\end{eqnarray}
with $H_\nu$ the well known Hermite polynomial of order $\nu$,
$E_\nu= (p_z^2+m^2+2\nu qB_m)^{1/2}$, the single particle energy eigen
value with $\nu=0,1,2,...$, the Landau quantum numbers and $L_y$, $L_z$
are length scales along $Y$ and $Z$ directions respectively.

In presence of strong quantizing magnetic field along Z-axis (as
specified by the choice of gauge) the momentum in the orthogonal plane
gets quantized and is given by $p_\perp=(2\nu qB_m)^{1/2}$, whereas, the
component along Z-axis varies continuously from $-\infty$ to $\infty$.
Further the phase space volume integral in momentum space in this
condition is given by
\begin{equation}
\frac{1}{(2\pi)^3}\int d^3p f(p)=
\frac{1}{(2\pi)^3}\int dp_z d^2p_\perp f(p)=\frac{qB_m}{4\pi^2}\sum_\nu
(2-\delta_{\nu 0})\int_{-p_f}^{+p_f} dp_z f(\nu,p_z)
\end{equation}
here we have assumed $c=\hbar=0$ and $p_f$ is the Fermi momentum.

To compute the Landau Fermi liquid interaction function $f$ from two
particle forward scattering matrix, we first derive from the first
principle the general form of two fermion transition matrix element. We
calculate for both scalar and vector coupling with direct and exchange
type interactions. Therefore, we have the two particle transition matrix
corresponding to scalar and vector boson exchange interactions
\begin{eqnarray}
T_{fi}&=&i\int j_{fi}^0(x)\phi(x)d^4x \\
T_{fi}&=&i\int j_{fi}^\mu (x)V^\mu(x)d^4x
\end{eqnarray}
where the fermion four-current is given by $j_{fi}^\mu(x)=g_k\bar
\psi_f(x)\gamma^\mu \psi_i(x)$ $\equiv (\rho(x), \vec j(x))$, 
with the spinors given by eqns.(3) and (4)
and $k=s$ and $k=v$ corresponding to scalar and vector coupling cases
respectively. The scalar and vector fields can very easily be obtained
from the well known Greens function solutions of Klein-Gordon
equations
\[
(\partial^2 +m^2)\phi,V^\mu =\rho, j^\mu
\]
and are given by
\begin{eqnarray}
\phi(x)&=&-\int \frac{d^4q}{(2\pi)^4} \frac{\exp[-iQ.(x-x')]}{Q^2-\zeta^2}
\rho(x')d^4x'\\
V^\mu(x)&=&-\int \frac{d^4q}{(2\pi)^4} \frac{\exp[-iQ.(x-x')]}{Q^2-\eta^2}
j^\mu(x')d^4x'
\end{eqnarray}
where $Q$ is the transferred four momentum ($Q^\mu \equiv
(q_0,q_x,q_y,q_z)$). 
Let us first compute the transition matrix element for scalar
interaction case. Substituting for the Fermi
spinors (eqns.(3) and (4)), we have
\begin{equation}
\rho(x')=\frac{g_s}{L_yL_z}\exp[-i\{(E_{\nu_1}- E_{\nu_2})
-(p_{1y}-k_{1y}) -(p_{1z}-k_{1z})\}][\rho]
\end{equation}
where
\begin{equation}
[\rho]=\frac{1}{2}\sum_{\rm{spin}} \bar u(x',\nu_2,k_1)\gamma^0
u(x',\nu_1,p_1)=\frac{1}{2}{\rm{Tr}}(\Lambda(x',\nu_1,\nu_2,p_1,k_1)
\gamma^0)
\end{equation}
Then we have after putting for $\rho(x')$ and integrating over
$t'$, $y'$ and $z'$, the direct part of
two-particle transition matrix element with scalar coupling interaction
\begin{eqnarray}
T_{fi}^{(d,s)}&=&-i\int \frac{d^4q}{(2\pi)}
\delta(q^0-E_{\nu_1}+E_{\nu_2})
\delta(q_y-p_{1y}+k_{1y})
\delta(q_z-p_{1z}+k_{1z}) \nonumber \\
&&\frac{\exp(-iQ.x)}{Q^2-\zeta^2}[\rho(x')]\frac{g_s}{L_yL_z} \exp(iq_xx')
dx' \frac{g_s}{L_yL_z}[\rho(x)]d^4x \nonumber \\
&&\exp[-i\{(E_{\nu_1^\prime}- E_{\nu_2^\prime})
-(p_{2y}-k_{2y}) -(p_{2z}-k_{2z})\}]
\end{eqnarray}
Integrating over $t$, $y$, $z$ and $q^0$, $q_y$, $q_z$, we get
\begin{equation}
T_{fi}^{(d,s)}=-i(2\pi)^3\delta(E)\delta(p_y)\delta(p_z)\frac{g_s^2}
{L_y^2L_z^2} \int \frac{dq_x}{2\pi}[\rho(x)][\rho(x')] \exp[-iq_x(x-x')]
\frac{1}{Q^2-\zeta^2}dxdx'
\end{equation}
where the $\delta$-functions indicate the abbreviated form of energy, 
$y$ and $z$ components of momentum conservation. In the derivation of eqn.(14) 
and the computation of subsequent matrix elements for direct case with vector 
boson exchange, we have assumed the two body process in the form
\begin{equation}
1 (p_1)+ 2 (p_2) \rightarrow 1^\prime (k_1)+ 2^\prime (k_2)
\end{equation}

In this case the  forward scattering amplitude can be obtained if we put
$E_{\nu_1}=E_{\nu_2}=E_\nu$ (say),
$E_{\nu_1^\prime}=E_{\nu_2^\prime}=E_{\nu^\prime}$ (say),
$p_{1y}=k_{1y}=p_y$ (say), $p_{2y}=k_{2y}=p_y^\prime$ (say),
$p_{1z}=k_{1z}=p_z$ (say) and  $p_{2z}=k_{2z}=p_z^\prime$
 (say). Then after evaluating the contour integral over $q_x$, given by
 \begin{equation}
I_{q_x}=\int_{-\infty}^{+\infty} \frac{dq_x}{2\pi} \frac{1}{Q^2-\zeta^2}
=-\frac{\exp[-\zeta\mid x-x'\mid]}{2\zeta}
\end{equation}
and substituting in two-body forward scattering matrix element given by 
eqn.(14), we have the Landau Fermi liquid interaction function corresponding 
to direct term for scalar coupling case
\begin{equation}
f_{\rm{Dir.}}^{(s)}=\frac{g_s^2}{8\zeta}\int \exp[-\zeta\mid
x-x'\mid][\rho(x)] [\rho(x')] dx dx'
\end{equation}
After evaluating the quantities within [~] (see Appendix A) we finally
get
\begin{eqnarray}
f_{\rm{Dir.}}^{(s)}&=&\frac{g_s^2}{32\zeta
E_\nu E_{\nu'}(E_\nu+m)(E_{\nu'}+m)}\int \exp[-\zeta\mid
x-x'\mid] dx  dx'\nonumber \\ && 
\{2E_\nu (E_\nu+m)(I^2_{\nu;p_y}(x)+I^2_{\nu-1;p_y}(x))
+2\nu qB_m (I^2_{\nu;p_y}(x)-I^2_{\nu-1;p_y}(x))\}\nonumber \\
&&\{2E_{\nu^\prime} (E_{\nu^\prime}+m)(I^2_{\nu^\prime;p_y^\prime}(x^\prime)+
I^2_{\nu^\prime-1;p_y^\prime}(x^\prime))
+2\nu^\prime qB_m (I^2_{\nu^\prime;p_y^\prime}(x^\prime)-
I^2_{\nu^\prime-1;p_y^\prime}(x^\prime)) \}
\end{eqnarray}
Then the quasi-particle energy is given by
\begin{eqnarray}
\varepsilon_\nu(p_z)&=&E_\nu(p_z)+\frac{1}{(2\pi)^2}
\sum_{\nu^\prime=0}^{[\nu_{\rm{max}}]}(2-\delta_{\nu'0})
\int_{x,x'=-\infty}^\infty \int_{p_{y'}=-\infty}^\infty
\int_{p_{z'}=-p_f}^{p_f} dx dx' dp_{y'} dp_{z'}
f_{\rm{Dir.}}^{(s)}(p,p';\nu, \nu')\nonumber \\
&=&E_\nu (p_z)+\Delta E_{\nu;{\rm{Dir.}}}^{(s)}(p_z)
\end{eqnarray}
where $[\nu_{\rm{max}}]=(\mu_f^2-m^2)/(2qB_m)$, the integer part of
right hand side.
In the case of ultra-strong magnetic field, $\nu=\nu'=0$, then
\begin{equation}
\varepsilon_0 (p_z)=E_0(p_z)+\frac{g_s^2}{\zeta^2}n_f
\end{equation}
where
\begin{equation}
n_f=\frac{qB_m}{2\pi^2}p_f
\end{equation}
the number density of fermions. The form of the result given in eqn.(20)
is identical with that of zero field case \cite{R18}. The energy density is 
then given by
\begin{equation}
\epsilon=\frac{qB_m}{4\pi^2}\sum_{\nu=0}^{[\nu_{\rm{max}}]} (2-\delta_{\nu
0})
\int_{-p_f}^{p_f} dp_z \varepsilon_\nu(p_z)
\end{equation}

Let us now compute the exchange part for scalar coupling case. In this
case we substitute $E_{\nu_1}=E_{\nu_2^\prime}=E_\nu$, $E_{\nu_1^\prime}=
E_{\nu_2}=E_{\nu^\prime}$, 
$p_{1y}=k_{2y}=p_y$, $p_{1z}=k_{2z}=p_z$, 
$p_{2y}=k_{1y}=p_y^\prime$, $p_{2z}=k_{1z}=p_z^\prime$,
to obtain the two-body forward scattering amplitude.
In this particular example, the basic process is 
\begin{equation}
1(p_1) +2(p_2) \rightarrow 1^\prime(k_2)+2^\prime(k_1)
\end{equation}
and the $q_x$ contour integral is given by
\begin{equation}
I_{q_x}=-\frac{\exp[-K\mid x -x'\mid]}{2K}
\end{equation}
where $K\approx (q_y^2+q_z^2+\zeta^2)^{1/2}$ and $q_i=(p_i-p_i^\prime)$,
with $i=y$ and $z$. Then evaluating $[\rho(x) \rho(x')]$ (see Appendix
A), we get
\begin{eqnarray}
f_{\rm{ex}}^{(s)}&=&\frac{g_s^2m}{32E_\nu E_{\nu^\prime}}\int dx dx' 
\frac{\exp[-K\mid x -x'\mid]}{K} \nonumber \\
&&[(E_\nu E_{\nu^\prime}+p_zp_z^\prime +m^2){\rm{Tr}}(AA^\prime)- \vec p_\perp.
\vec p_\perp^\prime {\rm{Tr}}(BB^\prime)]
\end{eqnarray}
where
\begin{equation}
{\rm{Tr}}(AA^\prime)=2(I_\nu(x)I_\nu(x')I_{\nu^\prime}(x)I_{\nu^\prime} (x')+
I_{\nu-1}(x)I_{\nu-1}(x')I_{\nu^\prime-1}(x)I_{\nu^\prime-1} (x'))
\end{equation}
and
\begin{equation}
{\rm{Tr}}(BB^\prime)=2(I_{\nu-1}(x)I_\nu(x')I_{\nu^\prime-1}(x)
I_{\nu^\prime} (x')+
I_\nu(x)I_{\nu-1}(x')I_{\nu^\prime}(x)I_{\nu^\prime-1} (x'))
\end{equation}
Then as before
\begin{eqnarray}
\Delta E_{\nu;{\rm{ex}}}^{(s)}&=& \frac{1}{(2\pi)^2}
\sum_{\nu^\prime=0}^{[\nu_{\rm{max}}]}(2-\delta_{\nu'0})
\int_{x,x'=-\infty}^\infty \int_{p_{y'}=-\infty}^\infty
\int_{p_{z'}=-p_f}^{p_f} dx dx' dp_{y'} dp_{z'}
f_{\rm{ex}}^{(s)}(p,p';\nu, \nu')
\end{eqnarray}
For $\nu=\nu^\prime=0$,
\begin{eqnarray}
\Delta E_{\nu;{\rm{ex}}}^{(s)}=
\frac{g_s^2}{8\pi^2}&&\int_{-p_f}^{p_f}dp_z^\prime \exp\left (
\frac{v^{\prime^2}}{qB_m}\right ) \frac{(E_0E_0^\prime+ p_zp_z^\prime
+m^2)}{E_0E_0^\prime}\nonumber \\ && \int_0^{\pi/2} {\rm{erfc}}\left (
\frac{v^\prime}{(2qB_m)^{1/2}}\sec\theta\right )\sec\theta d\theta
\end{eqnarray}
where $v^\prime \approx \mid p_z-p_z^\prime \mid $. In the zero field
case
\begin{equation}
f_{\rm{ex}}^{(s)}=\frac{g_s^2}{4E_0(p)E_0(p')} \frac{m^2-p.p'}
{(p'-p)^2+\zeta^2}
\end{equation}

We next consider the vector coupling case. The current four vector in the
direct case is given by
\begin{equation}
j^\mu(x)=g_v \frac{\exp[-i\{(E_{\nu_1}-E_{\nu_2})- (p_{1y}-
k_{1y}) -(p_{1z}-k_{1z})\}]}{L_yL_z} [j^\mu(x,p_1,k_1)]
\end{equation}
where 
\begin{equation}
[j^\mu]=\frac{1}{2}\sum_{\rm{spin}} \bar u \gamma^\mu u=\frac{1}{2}
{\rm{Tr}}(\Lambda(x,\nu_1,\nu_2,p_1,k_1)\gamma^\mu)
\end{equation}
Then it is very easy to show that the Landau Fermi liquid interaction
function is given by
\begin{equation}
f_{\rm{Dir}}^{(v)}=\frac{g_v^2}{32\eta E_\nu E_{\nu^\prime}}\int dx dx'
\exp[-\eta\mid x-x'\mid][(E_\nu  E_{\nu^\prime}-p_zp_{z'}){\rm{Tr}}A~
{\rm{Tr}}A'
+\vec p_\perp .\vec p_\perp^\prime {\rm{Tr}}B~ {\rm{Tr}}B']
\end{equation}
where we have used the same kind of substitutions as has been done for
the scalar case and further the relation,
\begin{equation}
{\rm{Tr}}(\Lambda(x,\nu,p)\gamma^\mu)=\frac{1}{2E_\nu}({\rm{Tr}}A
+{\rm{Tr}} B) p^\mu
\end{equation}
(see Appendix B) is used to obtain this expression.

The exchange term corresponding to vector coupling case is similarly
given by (see Appendix B for brief derivation)
\begin{equation}
f_{\rm{Ex}}^{(v)}=\frac{g_v^2}{32 E_\nu E_{\nu^\prime}}\int dx dx'
\frac{\exp[-K\mid x-x'\mid]}{K}
[(p_zp_{z'}- E_\nu  E_{\nu^\prime}+3m^2){\rm{Tr}}(AA')
-\vec p_\perp.\vec  p_\perp^\prime {\rm{Tr}}(BB')]
\end{equation}
 
 Then energy density of the system is given by (including all the terms) 
\begin{equation}
\epsilon =\frac{qB_m}{4\pi^2}\sum_{\nu=0}^{[\nu_{\rm{max}}]}
(2-\delta_{\nu 0})\left \{ \int_{-p_f}^{+p_f} dp_z E_\nu(p_z)
+\sum_{k,l}\int_{-p_f}^{+p_f} dp_z\Delta E_{\nu;(l)}^{(k)}(p_z)\right \}
\end{equation}
 where $k=s$ or $v$ and $l=$ Dir or ex. The kinetic pressure is then
 given by
 \begin{equation}
 P=\mu_f n_f -\epsilon
 \end{equation}
 The chemical potential or the Fermi energy is given by
\begin{eqnarray}
\mu_f=\varepsilon_\nu(p_z=p_f)&=&E_\nu(p_f)+\sum_{k,l}\Delta
E_{\nu;(l)}^{(k)} (p_f)\nonumber \\ &=& \mu_0 +\Delta \mu
\end{eqnarray}
Therefore knowing the modified form of single particle energy or the
quasi-particle energy, it is possible to obtain all the thermodynamic
quantities of the system from standard definitions. 
 
To obtain equation of state of dense neutron star matter or dense
electron gas, since both direct and exchange terms contribute in the case of 
$e+e \rightarrow e+e$ scattering, it is necessary to screen $e-e$ and
$e-p$ interactions 
in the direct case (in the case of $e-p$ scattering, of course, only the
direct term exists). To get the screening mass of electron, we proceed in
the following manner. We assume protons as the positively charged background. 
Then assuming local charge fluctuation we have
\begin{eqnarray}
n^+ &=& n_0^+ +n_1^+\nonumber \\
n_e&=&n_0^++n_e^\prime
\end{eqnarray}
In the case positive lattice background, we have to replace $n_0$ by
$Zn_0$, where $Z$ is the atomic number of the lattice ions. Then in the
case of proton background, the well known Poisson equation is given by
\begin{equation}
\nabla^2 \psi=-4\pi n_1^+ e+4\pi n_e^\prime e
\end{equation}
where $\psi$ is the electrostatic potential produced in the system
because of local charge fluctuation. Then we have from eqn.(21) after
substituting for $p_f$ in terms of chemical potential $\mu_f$ and
electrostatic potential $\psi$,
\begin{equation}
n_e=n_0^++\frac{eB_m}{2\pi^2}\sum_\nu (2-\delta_{\nu 0})\frac{\mu_fe\psi}
{(\mu_f^2-m^2-2\nu eB_m)^{1/2}}
\end{equation}
The last term is the perturbed part ($n_e^\prime$) as mentioned in
the second line of eqn.(39). Then we have 
\begin{equation}
\nabla^2 \psi=-4\pi n_1^+ e+\frac{2e^2B_m\mu_f}{\pi} \sum_\nu
(2-\delta_{\nu 0}) \frac{1}{(\mu_f^2-m^2-2\nu eB_m)^{1/2}} 
\end{equation}
which may be re-written in the form
\begin{equation}
\left [ \nabla^2 -k_{\rm{sc}}^2\right ]\psi=-4\pi n_1^+ e
\end{equation}
where $k_{\rm{sc}}$ is the screening mass and is given by
\begin{equation}
k_{\rm{sc}}= \left [ \frac{2e^2B_m\mu_f}{\pi} \sum_\nu
(2-\delta_{\nu 0}) \frac{1}{(\mu_f^2-m^2-2\nu eB_m)^{1/2}} \right ]^{1/2}
\end{equation}
and the screening length $r_D=1/k_{\rm{sc}}$ 

In the case of quark matter, however, only the exchange term contributes,
we therefore do not need quark-quark screening. Since colorless gluons do not
exist, the direct term has no significance.

\section{Conclusion}
In this article we have developed an exact formalism to obtain two-body
scattering matrix element for an wide range of magnetic field strength.
Although the result has been used to formulate an exact relativistic version 
of Landau theory of Fermi liquid, is also applicable to study weak and 
electromagnetic processes in presence of strong magnetic field. 

The formalism developed in this paper to compute Landau Fermi liquid
interaction $f$ can also be used to study magnetic properties of
fermionic system. The calculations (both theoretical and numerical) are of 
course quite complicated in the relativistic region (we shall report the result
in some future publication). The other interesting property- the transport 
theory of normal Fermi liquid can very easily be studied with the help of 
present formalism and obtain various kinetic coefficients along with their 
dependences on the strength of magnetic field.
\appendix
\section{}
In the direct case we need
\begin{equation}
\Lambda (\nu,x,x',k_y)=\sum_{\rm{spin}} 
u(\nu,x,k)\bar u(\nu,x',k)
\end{equation}
Substituting for the up and down spin solutions of Dirac equation
(eqns.(3) and (4)), we get
\begin{equation}
\Lambda=\frac{1}{2E_\nu}(Ak_\mu\gamma^\mu (\mu=0 ~~{\rm{and}}~~
z)\hfil\break+mA+B k_\mu\gamma^\mu (\mu=y ~~{\rm{and}}~~ p_y=p_\perp))
\end{equation}
The matrices $A$ and $B$ are given by
\begin{equation}
A=\left ( \begin{array}{l c c r}I_\nu I_\nu^\prime &0&0&0 \\
0 & I_{\nu-1}I_{\nu-1}^\prime &0 &0 \\
0 & 0 &I_\nu I_\nu^\prime &0 \\
0 & 0 & 0 &I_{\nu-1}I_{\nu-1}^\prime \\
\end{array}
\right )
\end{equation}
\begin{equation}
B= \left ( \begin{array}{l c c r}I_{\nu-1} I_\nu^\prime &0&0&0 \\
0 & I_\nu I_{\nu-1}^\prime &0 &0 \\
0 & 0 &I_{\nu-1} I_\nu^\prime &0 \\
0 & 0 & 0 &I_\nu I_{\nu-1}^\prime \\
\end{array}
\right )
\end{equation}
where the primes indicate the functions of $x'$.

Eqn.(A2) is an entirely new result and to the best of our knowledge it
has not been reported earlier. Further, this result plays the key role
in all kind of calculations related with electromagnetic and weak
interactions in presence of strong quantizing magnetic field.

Since $\gamma$ matrices are traceless and both $A$ and $B$ matrices are 
diagonal with identical blocks, it is very easy to evaluate the traces of 
the product of $\gamma$-matrices multiplied with any number of $A$ and/or $B$, 
from any side with any order e.g., 
\begin{equation}
{\rm{Tr}}(\gamma^\mu \gamma^\nu A_1A_2..B_1B_2..)=Tr(A_1A_2..B_1B_2..)
g^{\mu\nu},
\end{equation}
\begin{equation}
{\rm{Tr}}(\gamma^\mu\gamma^\nu\gamma^\lambda\gamma^\sigma
A_1A_2..B_1B_2..)=Tr(A_1A_2..B_1B_2..)(g^{\mu \nu}
g^{\sigma \lambda}-g^{\mu \lambda}g^{\nu \sigma}+g^{\mu \lambda}g^{\nu
\sigma}), 
\end{equation}
${\rm{Tr}}$(product of odd $\gamma$s with  $A$ and/or $B)=0$  etc. The other
interesting aspects of $A$ and $B$ matrices are:\\
i) $k_{1\mu}k^{2\mu}{\rm{Tr}}(A_1A_2)= (E_1E_2-k_{1z}k_{2z}){\rm{Tr}}(A_1A_2)$\\
ii) $k_{1\mu}k^{2\mu}{\rm{Tr}}(B_1B_2)= \vec k_{1\perp}.\vec
k_{2\perp}{\rm{Tr}}(B_1B_2)$\\
iii) $k_{1\mu}k^{2\mu}{\rm{Tr}}(A_1B_2)= k_{1\mu}k^{2\mu}{\rm{Tr}}(B_1A_2)=0$\\
iv) $p_{1\mu}k^{1\mu}p_{2\nu}k^{2\nu}{\rm{Tr}}(A_1B_2)\neq 0=
(E_{\nu_1}E_{\nu_2^\prime}-p_{1z}k_{1z})\vec p_{2\perp}.\vec k_{2\perp}
{\rm{Tr}}(A_1B_2)$
These set of relations are also totally new results and have not been
reported before.

To compute the direct part for scalar coupling case, we need
\begin{equation}
{\rm{Tr}}(\bar u(x,\nu,p)u(x,\nu,p)\gamma^0)=
{\rm{Tr}}(\Lambda(x,\nu,p) \gamma^0)
\end{equation}
Using the expression for $\Lambda$, it very easy to show that the above
trace is given by
\begin{equation}
\frac{1}{2E_\nu (E_\nu +m)} 
[2E_\nu (E_\nu +m)(I_{\nu;p_y}^2(x)+I_{\nu-1;p_y}^2(x))+2\nu
qB(I_{\nu;p_y}^2(x)-I_{\nu -1;p_y}^2(x))]
\end{equation}
To compute the exchange part for this case, we need
\begin{equation}
{\rm{Tr}}[(\bar u(x,\nu,p)\gamma^0u(\nu^\prime,x,p^\prime))
(\bar u(x',\nu^\prime,p^\prime)\gamma^0u(\nu,x',p))]
={\rm{Tr}}[\Lambda(x,x',p,\nu)\gamma^0\Lambda(x,x',p^\prime,\nu^\prime
)\gamma^0]
\end{equation}
Then by simple algebraic manipulation it is trivial to show that the
above trace is given by
\begin{equation}
\frac{1}{4E_\nu E_{\nu^\prime}}[(E_\nu E_{\nu^\prime}+p_zp_z^\prime
+m^2){\rm{Tr}}(AA') -\vec p_\perp .\vec p_\perp^\prime {\rm{Tr}}(BB')]
\end{equation}
where
\begin{equation}
{\rm{Tr}}(AA')=2(I_\nu(x)I_\nu(x')I_{\nu^\prime}(x)I_{\nu^\prime}(x')
+I_{\nu-1}(x)I_{\nu-1}(x')I_{\nu^\prime-1}(x)I_{\nu^\prime-1}(x'))
\end{equation}
\begin{equation}
{\rm{Tr}}(BB')=2(I_{\nu-1}(x)I_\nu(x')I_{\nu^\prime-1}(x)I_{\nu^\prime}(x')
+I_\nu(x)I_{\nu-1}(x')I_{\nu^\prime}(x)I_{\nu^\prime-1}(x'))
\end{equation}
\section{}
Now to compute two-particle forward scattering matrix element for vector
boson exchange case we need
\begin{equation}
{\rm{Tr}}(u(x,\nu,p)\bar u(x,\nu,p)\gamma^\mu)
\end{equation}
for direct case. It is trivial to show that the trace is given by
\begin{equation}
\frac{1}{2E_\nu}{\rm{Tr}}[(Ap_\sigma\gamma^\sigma +mA +Bp_\sigma
\gamma^\sigma)\gamma^\mu]=\frac{1}{2E_\nu}[p^\mu({\rm{Tr}} A +{\rm{Tr}}
B)]
\end{equation}
Then
\begin{equation}
[j^\mu(x)][j_\mu(x')]=\frac{1}{4E_\mu E_{\mu^\prime} }
[(E_\nu E_{\nu^\prime}-p_zp_z'){\rm{Tr}} A
{\rm{Tr}} A' +p_\perp p_\perp^\prime {\rm{Tr}} B {\rm{Tr}} B']
\end{equation}
In this case the terms ${\rm{Tr}}A {\rm{Tr}}B'$ and ${\rm{Tr}}A'
{\rm{Tr}}B$ do not contribute.
In the exchange interaction for vector coupling case, we need
\begin{equation}
{\rm{Tr}}[\Lambda(x,x',p,\nu)\gamma^\mu\Lambda(x,x',p^\prime,\nu^\prime
)\gamma^\mu]
\end{equation}
By some trivial algebra, we can very easily show that this expression is
given by
\begin{equation}
\frac{1}{4E_\nu E_{\nu^\prime}}{\rm{Tr}}[AA'p_\sigma p_\lambda^\prime
(\gamma^\sigma\gamma^\mu\gamma^\lambda\gamma_\mu)+ m^2AA'\gamma^\mu
\gamma_\mu+p_\sigma p_\lambda^\prime BB'(\gamma^\sigma\gamma^\mu\
\gamma^\lambda\gamma_\mu)]
\end{equation}
Then using the well known formula for the product of four
$\gamma$-matrices, we have the final form of the above expression
\begin{equation}
\frac{1}{4E_\nu E_{\nu^\prime}}[(p_zp_z^\prime -E_\nu E_{\nu^\prime})
{\rm{Tr}}(AA') +3m^2{\rm{Tr}}(AA')-\vec p_\perp . \vec p_\perp^\prime
{\rm{Tr}}(BB')]
\end{equation}
%------------------------------------------------------------


\begin{thebibliography}{99}
\bibitem{R1} R.C. Duncan and C. Thompson, Astrophys. J. Lett. {\bf{392}},
L9 (1992); C. Thompson and R.C. Duncan, Astrophys. J. {\bf{408}}, 194
(1993); C. Thompson and R.C. Duncan, MNRAS {\bf{275}}, 255 (1995);
C. Thompson and R.C. Duncan, Astrophys. J. {\bf{473}}, 322 (1996).
\bibitem{R2} P.M. Woods et al., Astrophys. J. Lett. {\bf{519}}, L139
(1999); 
 C. Kouveliotou, et al., Nature  {\bf{391}}, 235  (1999).
\bibitem{R3} K. Hurley, et al., Astrophys. Jour. {\bf{442}}, L111 (1999).
\bibitem{R4} S. Mereghetti and L. Stella, Astrophys. Jour. {\bf{442}},
L17 (1999);
J. van Paradihs, R.E. Taam and E.P.J. van den Heuvel,
Astron. Astrophys. {\bf{299}}, L41 (1995); S. Mereghetti,
astro-ph/99111252; see also A. Reisenegger, astro-ph/01003010.
\bibitem{R5} D. Bandopadhyaya, S. Chakrabarty, P. Dey
and S. Pal, Phys. Rev. {\bf{D58}}, 121301 (1998).
\bibitem{R6} S. Chakrabarty, D. Bandopadhyay and S. Pal, Phys. Rev.
Lett. {\bf{78}}, 2898 (1997);
D. Bandopadhyay, S. Chakrabarty and S. Pal, Phys. Rev.
Lett. {\bf{79}}, 2176 (1997).
\bibitem{R7} C.Y. Cardall, M. Prakash and J.M. Lattimer,
astro-ph/0011148 and references therein;
E. Roulet, Astro-ph/9711206; L.B. Leinson and A. P\'{e}rez, 
Astro-ph/9711216;
D.G. Yakovlev and A.D. Kaminkar, The Equation of States in
Astrophysics, eds. G. Chabrier and E. Schatzman P.214, Cambridge Univ.
\bibitem{R8}S. Chakrabarty and P.K. Sahu, Phys. Rev. {\bf{D53}}, 4687 (1996).
\bibitem{R9}S. Chakrabarty, Phys. Rev. {\bf{D51}}, 4591 (1995);
Chakrabarty, Phys. Rev. {\bf{D54}}, 1306 (1996).
\bibitem{R10} T. Ghosh and S. Chakrabarty, Phys. Rev. {\bf{D63}},
0403006 (2001); T. Ghosh and S. Chakrabarty, Int. Jour. Mod. Phys.
{\bf{D10}}, 89 (2001).
\bibitem{R11} V.G. Bezchastrov and P. haensel, Astro-ph/9608090.
\bibitem{R12} S. Mandal and S. Chakrabarty (two papers on weak and
electromagnetic processes to be published).
\bibitem{R13} A. Melatos, Astrophys. Jour. {\bf{519}}, L77 (1999);
A. Melatos, MNRAS {\bf{313}}, 217 (2000).
\bibitem{R14} S. Bonazzola et al, Astron. \& Atrophysics. {\bf{278}},
421 (1993); M. Bocquet et al, Astron. \& Atrophysics. {\bf{301}}, 757
(1995).
\bibitem{R15} B. Bertotti and A.M. Anile, Astron. \& Atrophysics.
{\bf{28}}, 429 (1973); C. Cutler and D.I. Jones, gr-qc/0008021;
 K. Konno, T. Obata and Y. Kojima, gr-qc/9910038; A.P.
Martinez et al, hep-ph/0011399;
 M. Chaichian et al, Phys. Rev. Lett. {\bf{84}}, 5261
(2000); Guangjun Mao, Akira Iwamoto and Zhuxia Li, astro-ph/0109221;
 A. Melatos, Astrophys. Jour. {\bf{519}}, L77 (1999);
A. Melatos, MNRAS {\bf{313}}, 217 (2000); R. Gonz\'{a}lez Felipe et al,
astro-ph/0207150 and references therein.
\bibitem{R16} S. Mandal and S. Chakrabarty (submitted).
\bibitem{R17} Issac Vida\~{n}a and Ignazio Bombaci, nucl-th/0203061.
\bibitem{R18} G. Baym and S.A. Chin, Nucl. Phys. {\bf{A262}}, 527
(1976).
\end{thebibliography}
\end{document}